\documentclass[hyper,a4paper,12pt]{JHEP3}
\usepackage{graphicx}
\usepackage{epsfig}
\usepackage{amsmath}
\usepackage{amssymb}
\usepackage{amsthm}

\usepackage{ulem}
\def\unit{{1\kern-.65ex {\rm l}}}
\def\1{{1\kern-.65ex {\rm l}}}

\def\Im{\mathop{\mathrm{Im}}\nolimits}

\let\vev=\bracket

\title{Effective Mass of Holographic Brownian Particle in Rotating Plasma}
\author{Ardian Nata Atmaja$^{1}$\\
${}^1$Research Center for Physics, Indonesian Institute of Sciences (LIPI)\\
~Kompleks PUSPITEK Serpong, Tangerang 15310, Indonesia.\\
email: \email{ardian@teori.fisika.lipi.go.id}
}

\abstract{The dynamic of string fluctuation under rotating BTZ black hole is studied using the method of~\cite{Son:2009vu}. We compare the result with previous computation in~\cite{Atmaja:2012jg}, with a different method, for the case of co-rotating string. The result gives definite answer that the end of string identified as an external quark at the boundary behaves as Brownian particle with mass is given an effective mass $M_{eff}$ equal to the zero temperature mass of external quark $M_0$ times Lorentz factor $\gamma$ to the power three. We also extend the computation for general case of rotating black hole where the metric is asymptotically $AdS_3$ and the fluctuation is taken only along the co-rotating motion. It shows that the effective mass of the external quark is in general given by $\gamma^3 M_0$.}

\begin{document}

\section{Introduction}

Brownian motion is a phenomena where a Brownian particle, refers to an external quark in our case, moves randomly in a thermal plasma medium as the consequence of random collisions with the medium constituent. In a standard formula, the dynamic of Brownian particle is described by the Langevin equation
\begin{align}
 m{d^2x \over dt^2}=-\mu~{dx\over dt}+R(t),\ \ \ \ \vev{R(t)R(t')}=2\mu T\delta(t-t'),
\end{align}
where $m$ is the mass of Brownian particle, $\mu$ is the friction coefficient, $R(t)$ is the random force, and $T$ is temperature of the medium. The mass of Brownian particle may not be equal to the rest mass of the external quark and it might be modified due to the medium where it moves. Knowing temperature of the media and by computing the early and late times behavior of the displacement square of the Brownian particle, we can extract the mass and friction coefficient respectively~\cite{Uhlenbeck:1930zz, deBoer:2008gu}.

In strongly coupled medium such as Quark Gluon Plasma, the calculation for random-random force corellator must rely on non-perturbative techniques. An elegant technique that we are going to present here is by using the holographic principal in the context of AdS/CFT correspondence~\cite{Witten:1998qj,Gubser:1998bc}. The setting in this holographic Brownian motion is given by a fundamental string stretching from horizon to the boundary of a black hole background metric. The Brownian motion corresponds to fluctuation of the end of fundamental string at the boundary, which is interpreted as the external quark~\cite{Maldacena:1998im,Rey:1998ik}, and the thermal medium corresponds to the black hole metric.

There are two methods of computing holographic Brownian motion. The first method~\cite{deBoer:2008gu} considers Brownian motion of the string at the boundary caused by Hawking radiation near the horizon while the second method~\cite{Son:2009vu} uses the extension of Schwinger-Keldysh formalism for thermal field theory at the boundary. Although both methods use different approach, they surprisingly give the same physical result of Brownian motion at the boundary even though they have their own eminences yet the connection between these methods is unclear, for details see~\cite{Hubeny:2010ry}. As an example, the first method can be used to calculate Brownian motion where the background metric is not necessarily asymptotically Anti de-Sitter spacetime however the calculations must rely on the presumable Langevin equation~\cite{Tong:2012nf,Edalati:2012tc}. On the other hand the second method requires the background metric to be asymptotically Anti de-Sitter but no a priori Langevin equation is taken into account. The Langevin equation comes out naturally from the path integral integration which can be an advantage if we would like to calculate Brownian motion where the exact form of the Langevin equation is unknown, e.q. in rotating fluid or nonlinear Langevin equation.
  
This article is motivated by an interesting result of the holographic Brownian motion in two dimensional rotating plasma computed in~\cite{Atmaja:2012jg}. Using the first method, it was concluded that the mass of the co-rotating Brownian particle appears in the Langevin equation should be modified into an effective mass rather than being zero temperature mass of an external quark $M_Q^0$. The modification on the mass of Brownian particle was also suggested in the non-rotating case for which the effective mass is the sum of the zero temperature quark mass and the in-medium mass shift $\Delta M$ effected by the medium~\cite{Son:2009vu}\footnote{This in-medium mass shift was first calculated in the drag force computation~\cite{Herzog:2006gh}.}. On contrary the effective mass in co-rotating case of~\cite{Atmaja:2012jg} did not have in-medium mass shift term instead it was given by $M_Q^0$ times a multiplication factor $\gamma^3$, where $\gamma$ is a ``Lorentz'' factor. The arguments used in~\cite{Atmaja:2012jg} was based on observation of asymptotic time behaviour of the displacement squared and it was supported by a new approach using random-random force correlator. Nevertheless, It might still raise some doubts such as whether the Langevin equation used in~\cite{Atmaja:2012jg} was correct or not. In order to address this question a method describing holographic Brownian motion formulated by Son and Teaney~\cite{Son:2009vu} is of the interest. 

In this article we would like to redo the calculation in~\cite{Atmaja:2012jg} for BTZ black hole case using the second method and compare it with the previous calculation. We also generalize the calculation for arbitrary rotating black holes, with one rotation parameter, where the metric is taken to be asymptotically Anti de-Sitter and restrict to the Brownian motion of co-rotating string in this background\footnote{Here we only consider the fluctuation direction along the string velocity.}. The situation here is some what different from relativistic Brownian motion as in~\cite{Giecold:2009cg} though the fluctuation is carried out over a moving string\footnote{The non co-rotating string case is in principal similar to the relativistic Brownian motion.}. We will check if this ``Lorentz'' factor in the effective mass is generic or not.

\section{Brownian Motion in BTZ Black Hole}
\subsection{Co-rotating solution}

First, we start with the dynamic of a string which is classically given by the Nambu-Goto action
\begin{eqnarray}
\label{NG action}
 S_{NG}&=&-{1 \over 2\pi\alpha'}\int d\sigma^2 \sqrt{(\dot{X}. X')^2-\dot{X}^2 X'^2},
\end{eqnarray}
with
\begin{eqnarray}
 \dot{X}. X'&=&g_{\mu\nu}{dX^\mu \over d\sigma^0}{dX^\nu \over d\sigma^1},\\
\dot{X}^2&=&g_{\mu\nu}{dX^\mu \over d\sigma^0}{dX^\nu \over d\sigma^0},\\
X'^2&=&g_{\mu\nu}{dX^\mu \over d\sigma^1}{dX^\nu \over d\sigma^1},
\end{eqnarray}
where $\sigma\equiv(\sigma^0,\sigma^1)$ is the string worldsheet coordinate, $X^\mu$ is the spacetime coordinate, $\alpha'$ is related to the string length scale, and $g_{\mu\nu}$ is the background metric. 

The background metric, $g_{\mu\nu}$, for BTZ black hole is written as follows~\cite{Banados:1992wn}:
\begin{align}
\label{BTZ metric}
 ds^2&=-N^2dt^2+\frac{1}{N^2}dr^2+r^2\left(N^\phi dt+d\phi\right)^2,\nonumber\\
N^2(r)&=-M+\frac{r^2}{l^2}+\frac{J^2}{4r^2},\ \ \ \ \ N^\phi (r)=-\frac{J}{2r^2},
\end{align}
with $-\infty<t<\infty,~0<r<\infty$ and $0\leq\phi< 2\pi$. $M$ and $J$ are mass and angular momentum of the black hole. The Hawking temperature for this metric, also interpreted as temperature of the plasma, is given by a formula
\begin{align}
\label{temperature}
 T_H&=\frac{r_+^2-r_-^2}{2\pi l^2 r_+},\nonumber\\
 r_{\pm}=\sqrt{\frac{Ml^2}{2}\pm\frac{1}{2}\xi},&\ \ \ \ \xi=\sqrt{M^2l^4-J^2l^2},
\end{align}
where $r_-$ is the inner horizon and $r_+$ is the outer horizon. For convenience, we will set the length square dimension $l^2=1$.

The string equation of motion can be derived from the Nambu-Goto action with a linear ansatz
\begin{align}
 \phi(t,r)=w t+\eta(r),
\end{align}
where $w$ is a constant velocity of the string and $\eta$ is an effective field which depend only on the radial coordinate. Physical solution requires
\begin{align}
 \eta'(r)=-\frac{\pi_\phi}{r^2N^2}\sqrt{\frac{r^4\left(w+N^\phi\right)^2-r^2N^2}{\pi_\phi-r^2N^2}},
\end{align}
where $\pi_\phi$ is proportional to the energy flow along the string which is fixed to be $\pi_\phi^2=r_{NH}^2 N^2(r_{NH})$, with $r^2_{NH}={r_+^2+r_-^2-2r_+r_- w \over 1-w^2}$. The co-rotating string in~\cite{Atmaja:2012jg} was defined at $\pi_\phi=0$ or equal to set the velocity $w=r_-/r_+$ and $\eta'(r)=0$.

\subsection{Small fluctuation}
Following~\cite{Atmaja:2012jg}, we take small fluctuation around co-rotating solution
\begin{align}
 \phi(t,r)=w t+C,
\end{align}
with $w=r_-/r_+$ and $C$ is constant, which can be set to zero. Writing $\phi\to w t+ \phi$  we have
\begin{eqnarray}
 \dot{X}. X'&=&r^2\phi'\left(N^\phi+w+\dot{\phi}\right),\\
\dot{X}^2&=&-N^2+r^2\left(N^\phi+w+\dot{\phi}\right)^2,\\
X'^2&=&{1\over N^2}+r^2\phi'^2,
\end{eqnarray}
so the determinant of the metric becomes
\begin{align}
 -g=1+r^2N^2\phi'^2-{r^2\over N^2}\left(N^\phi+w+\dot{\phi}\right)^2.
\end{align}
Expanding the Nambu-Goto action up to quadratic power of $\phi$ we obtain equation of motion
\begin{align}
\label{EOM linear}
 -\frac{r^2N}{\left(N^2-r^2(w+N^\phi)^2\right)^{3/2}}\ddot{\phi}+\frac{\partial}{\partial r}\left[\frac{r^2N^3\phi'}{\left(N^2-r^2(w+N^\phi)^2\right)^{1/2}}\right]=0.
\end{align}
The equation of motion could also be derived by considering the small fluctuation around static string solution over a background metric
\begin{align}
\label{metric static}
 ds^2=&-\sqrt{N^2(N^2-r^2(w+N^\phi)^2)}dt^2+{1\over \sqrt{N^2(N^2-r^2(w+N^\phi)^2)}} dr^2\nonumber\\
     &+{r^2N^2 \over N^2-r^2(w+N^\phi)^2}d\phi^2.
\end{align}
This metric was used in~\cite{Atmaja:2012jg} in the computation of the Brownian motion. Unfortunately this metric does not manifest asymptotically AdS near the boundary, $r\to\infty$.
There is actually a nicer metric implying the same equation of motion with asymptotically AdS near the boundary and it is given by\footnote{Though the metric (\ref{metric static}) is different from (\ref{nicer metric}), the computation in~\cite{Atmaja:2012jg} gives the same result since the near horizon structure and the $g_{\phi\phi}$ component of both metrics are equal.}
\begin{align}
\label{nicer metric}
 ds^2=&-(N^2-r^2(w+N^\phi)^2)dt^2+{1\over N^2} dr^2+{r^2N^2 \over N^2-r^2(w+N^\phi)^2}d\phi^2.
\end{align}
For our purpose here we are going to use the metric (\ref{nicer metric}) as such to follow the computation in~\cite{Son:2009vu} for asymptotically AdS metric background. In doing so, we transform the time and angular coordinates to
\begin{align}
 t=\gamma \bar{t},\ \ \ \ \phi={1\over\gamma} \bar{\phi},\ \ \ \ \gamma={r_+ \over \sqrt{r_+^2-r_-^2}}={1\over\sqrt{1-w^2}},
\end{align}
where $\gamma$ is the ``Lorentz'' factor. In this coordinate, the Hawking temperature changes to
\begin{align}
 \bar{T}_H={\sqrt{r_+^2-r_-^2}\over 2\pi}=\gamma T_H\equiv \bar{T}
\end{align}
and the equation of motion is modified to
\begin{align}
\label{new EOM linear}
 -\frac{r^2N}{\left(N^2-r^2(w+N^\phi)^2\right)^{3/2}}{\ddot{\bar{\phi}}\over \gamma^2}+\frac{\partial}{\partial r}\left[\frac{r^2N^3\bar{\phi}'}{\left(N^2-r^2(w+N^\phi)^2\right)^{1/2}}\right]=0,
\end{align}
with $\dot{}\equiv{\partial \over \partial_{\bar{t}}}$. The solution is written as\footnote{One can derive the relation between $f$ and $\bar f$ from equation (\ref{new EOM linear}) or from $\phi(t)={1\over\gamma}\bar\phi(\bar{t})$.}
\begin{align}
 \bar{\phi}=e^{-i\bar{\omega}\bar{t}}\bar{f}_{\bar{\omega}}(r),\ \ \ \ \bar{f}_{\bar{\omega}}(r)= f_{(\bar{\omega}/\gamma)}(r),
\end{align}
with two solutions~\cite{Atmaja:2012jg}
\begin{align}
 f^{\pm}_{\omega}(r)=\frac{\left(1\pm\frac{2i\chi}{\sqrt{s+\xi}}\right)}{\left(1+\sqrt{1+\frac{s}{\xi}}\right)^{\pm2\frac{i\chi}{\sqrt{\xi}}}}{\left(s\over \xi\right)}^{\pm\frac{i\chi}{\sqrt{\xi}}},\ \ \ \ s=r^2-r_+^2,\ \ \ \ \chi=\gamma{\omega\over 2}.
\end{align}
We have normalized the solutions at the boundary, $f^{\pm}_{\omega}(r\to\infty)=1$. 

\subsection{Langevin equation}
Imposing another boundary condition near horizon for the full Kruskal plane as shown in ~\cite{Son:2009vu}, with right and left quadrant denoted by subscript 1 and 2 respectively, the general solution can be written as
\begin{align}
 \bar{\phi}_1(\bar{\omega},r)&=a(\bar{\omega})\bar{f}^-_{\bar{\omega}}(r)+b(\bar{\omega})\bar{f}^+_{\bar{\omega}}(r),\\
 \bar{\phi}_2(\bar{\omega},r)&=a(\bar{\omega})e^{-\bar{\omega}\sigma}\bar{f}^-_{\bar{\omega}}(r)+b(\bar{\omega})e^{-\bar{\omega}(\sigma-1/ \bar{T})}\bar{f}^+_{\bar{\omega}}(r),
\end{align}
with $0\leq\sigma\leq 1/\bar{T}$ and
\begin{align}
 a(\bar{\omega})&=\bar\phi_1(\bar\omega)(1+n(\bar\omega))-\bar\phi_2(\bar\omega)e^{\bar\omega\sigma}n(\bar\omega)\\
 b(\bar{\omega})&=\bar\phi_2(\bar\omega)e^{\bar\omega\sigma}n(\bar\omega)-\bar\phi_1(\bar\omega)n(\bar\omega).
\end{align}
The Boltzmann-Einstein distribution is given by $n(\bar\omega)=1/(e^{\bar\omega/\bar{T}}-1)$. Notice that $\bar{\phi}_i(\bar{\omega},r\to\infty)=\bar{\phi}_i({\bar\omega})$, with $i=1,2$. Plugging back the solution into the action resulting the boundary action
\begin{align}
 S_{bdr}&=S_1-S_2\notag\\
&=-{T'\over 2}\int{d\bar\omega\over 2\pi}\lim_{r\to r_b}\bar\phi_1(-\bar\omega,r)\partial_r \bar\phi_1(\bar\omega,r)-(1\to 2),\\
 T'&= {r_b^4\over 2\pi\alpha'},
\end{align}
where we have put a UV-cutoff at $r=r_b\gg r_+$ and brackets in the second term is the first term with index 1 replaced by index 2. Writing in terms of the bulk $r$ and $a$ fields
\begin{align}
 \bar\phi_r(\bar\omega,r)={1\over 2}\left(\bar\phi_1(\bar\omega,r)+\bar\phi_2(\bar\omega,r)\right),\ \ \ \ \bar\phi_a(\bar\omega,r)=\bar\phi_1(\bar\omega,r)-\bar\phi_2(\bar\omega,r),
\end{align}
and defining a retarded Green function
\begin{align}
G^0_R(\bar\omega)=T' \bar{f}^+_{\bar\omega}(r_b)\partial_r \bar{f}^-_{\bar\omega}(r_b), 
\end{align}
the boundary action becomes, for $\sigma=0$,
\begin{align}
 S_{bdr}&=\int{d\bar\omega\over 2\pi}\left[-\bar\phi_a(-\bar\omega)G^0_R \bar\phi_r(\bar\omega)+{i\over 2}\bar\phi_a(-\bar\omega)G_{sym} \bar\phi_a(\bar\omega)\right],
\end{align}
where $G_{sym}(\bar\omega)=-(1+2n(\bar\omega)) \Im{G^0_R(\bar\omega)}$. With this boundary action we can extract the Langevin equation from the partition function, see~\cite{Son:2009vu} for details,
\begin{align}
\label{LE in momentum}
 [-M_0 \bar\omega^2+G_R(\bar\omega)]\bar\phi_r(\bar\omega)=\bar\xi(\bar\omega),\ \ \ \ \vev{\bar\xi(\bar\omega)\bar\xi(\bar\omega)}=G_{sym}(\bar\omega),
\end{align}
with $G_R(\bar\omega)=G^0_R(\bar\omega)+M_0 \bar\omega^2$ and $M_0$ is the zero temperature mass of external quark,
\begin{align}
 M_0={r_b\over 2\pi\alpha'},
\end{align}
computed from the asymptotically AdS metric (\ref{nicer metric}) which turns out to be equal to the zero temperature mass of external quark in BTZ black hole. In expansion of $G^0_R(\bar\omega)$ for large $r_b$ up to zero order of $r_b$, we then obtain
\begin{align}
 G_R(\bar\omega)=-{i\over 2\pi\alpha'}\left({r_+^2\over\gamma^2}+\bar\omega^2\right)\bar\omega.
\end{align}
The result shows there is no in-medium mass shift coming from the fluid which is dual to the metric (\ref{nicer metric}). Taking low frequency limit of $G_R(\bar\omega)$, up to $\bar\omega^2$, the Langevin equation becomes
\begin{align}
 \left[-{M_0} \bar\omega^2-i{r_+^2\over 2\pi\alpha'\gamma^2}\bar\omega\right]\bar\phi_r(\bar\omega)=\bar\xi(\bar\omega),\ \ \ \ \vev{\bar\xi(-\bar\omega)\bar\xi(\bar\omega)}=2\bar{T}{r_+^2\over 2\pi\alpha'\gamma^2}.
\end{align}
Defining the boundary field in the ``original'' time coordinate $t$ with $\phi(t)=\int {d\omega\over 2\pi}e^{-i\omega t}\bar\phi({\omega\gamma})$, where we have suppressed the subscript index $r$, the Langevin equation in time coordinate is
\begin{align}
\label{BTZ result}
 {M_0\gamma^3} {\partial^2 \phi(t) \over \partial t^2}+{r_+^2\over 2\pi\alpha'}{\partial \phi(t) \over \partial t}={\xi(t)},\ \ \ \ \vev{\xi(t)\xi(t')}=2T{r_+^2\over 2\pi\alpha'}\delta(t-t'),
\end{align}
where we have defined $\xi(t)=\bar\xi(\bar{t})$.
Here the effective mass comes out naturally and it turns out to be equal to effective mass proposed in~\cite{Atmaja:2012jg}. The result also justifies the Langevin equation and the random-random force correlator approach used in~\cite{Atmaja:2012jg}.

\section{General Co-rotating Brownian Motion in Rotating Fluid}

The most common metric components for rotating AdS black holes in more than three dimensional spacetime depend on more than one coordinate: a radial coordinate and some compact coordinates~\cite{Klemm:1997ea,Hawking:1998kw,Gibbons:2004uw}. We limit our case up to five dimensional spacetime, where the metric components depend on only one compact coordinate, and take the metric components to only depend on radial coordinate by means of fixing the compact coordinate in such away the metric of the subspacetime is asymptotically AdS and also it has stationary string solution. For example in four dimensional Kerr-AdS black holes, after converting to asymptotically ``cannonical'' AdS coordinate~\cite{Gibbons:2004ai}, one can take the polar coordinate $\hat\theta=\pi/2$ so that the metric is asymptotically $AdS_3$. Note that this way of subspacetime construction may not be a solution to the Einstein equation with negative cosmological constant and asymptotically the coordinate only covers the half-space of the $AdS_3$ space. As an example, a simple way of getting asymptotically $AdS_3$ out of four dimensional toroidal KMV-AdS black hole in~\cite{Klemm:1997ea} is by setting $P=0$. The resulting metric is clearly not the solution to three dimensional Einstein equation with negative cosmological constant.

Following the above construction, we may consider the general metric with one rotation in $X$ direction
\begin{align}
\label{general metric}
 %ds^2=-h_t f dt^2+ {h_r\over f} dr^2+g_x\left(dX+g_t dt\right)^2
  ds^2=-h_t f dt^2+ {h_r\over f} dr^2+g_x\left(dX+g_t dt\right)^2.
\end{align}
Asymptotically AdS requires
\begin{align}
 h_t f\sim r^2, h_r/f\sim r^{-2}, g_x\sim r^2,g_t\sim O(r^{-2}).
\end{align}
As mentioned before, this metric may not be solution of the three dimensional Einstein equation with negative cosmological constant. The near boundary expansion for $g_t-$component is different compare to the expansion in~\cite{Henneaux:1985tv} with $O(r^{-3})$. Nevertheless, the asymptotically AdS behaviour above cover all possible subspacetime of the more than three dimensional spacetime and also including the BTZ black hole (\ref{BTZ metric}). More precisely, we also consider the general metric to be a black hole with an even horizon at $r_+$. All functions in the metric components are positive semidefinite and regular in $r_+\le r<\infty$ with an exception $f\sim (r-r_+)$ near the horizon.

\subsection{Co-rotating solution}
 The dynamic of fundamental string described by Nambu-Goto action under the general metric above with linear solution of the equation of motion is given by
\begin{align}
 X(t,r)=v t+\eta(r),
\end{align}
where $v$ is constant and here we have used the axial gauge. The curved function
\begin{align}
 \eta'(r)=-{\pi_X\over g_xf}\sqrt{h_r \over h_t}\sqrt{g_x h_t f-g_x^2 (v+g_t)^2 \over g_x h_t f-\pi_X^2}
\end{align}
is related to a drag force, where $\pi_X$ is the total force acts on the string~\cite{Gubser:2006bz,Herzog:2006gh}. Reality condition requires that $\pi_X=g_x(v+g_t)|_{r_{c}}$ where $r_c\ge r_+$ is the critical point defined as the largest positive root of equation $h_t f=g_x(v+g_t)^2$. The co-rotating solution is defined as the solution for $\pi_X=0$ thus $g_t(r_c)=-v$, here $r_c=r_+$, and so $\eta=constant$ which we set to be zero~\cite{NataAtmaja:2010hd, Atmaja:2012jg}. 

\subsection{Small fluctuation}
Taking small fluctuation around this co-rotating solution, the metric determinant in the square root of Nambu-Goto action (\ref{NG action}) becomes
\begin{align}
 -g=h_th_r+g_x h_t f X'^2-g_x{h_r\over f}\left(\dot{X}+g_t-g_t(r_c)\right)^2,
\end{align}
with $\dot{X},X'\ll$ is small. The Nambu-Goto action up to the second order expansion of small fluctuation is then written as
\begin{align}
 S_{NG}&\approx S_{(0)}+S_{(1)}+S_{(2)}+\cdots.\\
 S_{(0)}&=-T_s\int d\sigma^2 \sqrt{-g_0}\\
\label{NG expansion}
 S_{(1)}&=T_s\int d\sigma^2 {(g_t-g_t(r_c))g_x h_r \over f\sqrt{-g_0}}\dot{X}\\
 S_{(2)}&=-{T_s\over 2}\int d\sigma^2 {g_x h_t h_r \over (-g_0)}\left[{f\sqrt{-g_0}\over h_r}X'^2-{h_r\over f\sqrt{-g_0}}\dot{X}^2\right],
\end{align}
with
\begin{align}
 -g_0=h_th_r-g_x{h_r\over f}\left(g_t-g_t(r_c)\right)^2,\ \ \ \ T_s={1\over 2\pi\alpha'}.
\end{align}
The above expansion of Nambu-Goto action can be obtain effectively from small fluctuation over a static solution of the following effective metric\footnote{The expansion using the effective metric does not contain odd orders of the small fluctuation $X$. However, since at the end we only look for the on-shell action upto second order expansion the odd term (\ref{NG expansion}) in the original Nambu-Goto expansion does not contribute to the calculation.}
\begin{align}
 ds^2=-{f\over h_r}(-g_0)dt^2+{h_r\over f}dr^2+{g_x h_t h_r \over (-g_0)}dX^2.
\end{align}
The Hawking temperature in this effective metric is given by
\begin{align}
 T_H=T={f'(r_c)\over 4\pi}\sqrt{h_t(r_c)\over h_r(r_c)}.
\end{align}
Recall that for co-rotating motion $r_c=r_+$ which is the even horizon of the original rotating metric.

Near the boundary, $r\to\infty$, the metric components behave as
\begin{align}
 g_{tt}\sim -r^2(1-v^2),\ \ \ g_{rr}\sim r^{-2},\ \ \ g_{xx}\sim r^2 (1-v^2)^{-1}.
\end{align}
We can write it in form of the usual asymptotically AdS metric in terms of ``bar'' coordinates
\begin{align}
  t=\gamma\bar{t},\ \ \ \ X={1\over\gamma} \bar{X},\ \ \ \ \gamma={1\over\sqrt{1-v^2}}.
\end{align}
In this ``bar'' coordinates, the Hawking temperatures is scaled as
\begin{align}
 \bar{T}= \gamma T.
\end{align}
Equation of motion for $\bar{X}$ is
\begin{align}
 -{g_x h_t h_r^2 \over f (-g_0)^{3/2}}{1\over\gamma^3}\ddot{\bar{X}}+{\partial\over\partial_r}\left({g_x h_t f\over \gamma\sqrt{-g_0}}\bar{X}'\right)=0.
\end{align}
The solution is written as $\bar{X}(\bar\omega,r)=e^{-i\bar\omega\bar{t}}F_{\bar\omega}(r)$. Two independent solutions $F^+_{\bar\omega}$ and $F^-_{\bar\omega}$ denoting the outgoing and incoming wave functions near the horizon respectively. These solutions are related by $F^-_{\bar\omega}=(F^+_{\bar\omega})^*$.

We set up boundary condition at the boundary such that $F^\pm_{\bar\omega}(r\to\infty)=1$. In order to have a thermal field theory as the boundary theory, we have to impose another boundary condition near horizon, see~\cite{Son:2009vu} for detail procedure. 

\subsection{Langevin equation}
In the full Kruskal plane with right and left quadrant denoted by subscript 1 and 2 respectively, the general solution can be written as
\begin{align}
 \bar{X}_1(\bar{\omega},r)&=a(\bar{\omega})F^-_{\bar{\omega}}(r)+b(\bar{\omega})F^+_{\bar{\omega}}(r),\\
 \bar{X}_2(\bar{\omega},r)&=a(\bar{\omega})e^{-\bar{\omega}\sigma}F^-_{\bar{\omega}}(r)+b(\bar{\omega})e^{-\bar{\omega}(\sigma-1/ \bar{T})}F^+_{\bar{\omega}}(r),
\end{align}
with $0\leq\sigma\leq 1/\bar{T}$ and
\begin{align}
 a(\bar{\omega})&=\bar{X}_1(\bar\omega)(1+n(\bar\omega))-\bar{X}_2(\bar\omega)e^{\bar\omega\sigma}n(\bar\omega)\\
 b(\bar{\omega})&=\bar{X}_2(\bar\omega)e^{\bar\omega\sigma}n(\bar\omega)-\bar{X}_1(\bar\omega)n(\bar\omega).
\end{align}
The Boltzmann-Einstein distribution is given by $n(\bar\omega)=1/(e^{\bar\omega/\bar{T}}-1)$ and $\sigma$ parameterizes the vertical distance between time in quadrant 1 and 2 in the complex time diagram, for our purpose here we set $\sigma=0$. Plugging back the solution into the action gives a boundary action
\begin{align}
 S_{bdr}&=-{T'\over 2}\int{d\bar\omega\over 2\pi}\lim_{r\to r_b}\bar{X}_1(-\bar\omega,r)\partial_r \bar{X}_1(\bar\omega,r)-(1\to 2),\\
 T'&= {r_b^4\over 2\pi\alpha'},\\
\end{align}
where we have put a UV-cutoff at $r=r_b\gg r_c$ and brackets in the second term is the first term with index 1 replaced by index 2. In terms of the bulk $r-$ and $a-$fields,
\begin{align}
 \bar{X}_r(\bar\omega,r)={1\over 2}\left(\bar{X}_1(\bar\omega,r)+\bar{X}_2(\bar\omega,r)\right),\ \ \ \ \bar{X}_a(\bar\omega,r)=\bar{X}_1(\bar\omega,r)-\bar{X}_2(\bar\omega,r),
\end{align}
and defining a retarded Green function
\begin{align}
G^0_R(\bar\omega)=T' F^+_{\bar\omega}(r_b)\partial_r F^-_{\bar\omega}(r_b), 
\end{align}
the boundary action becomes
\begin{align}
 S_{bdr}&=\int{d\bar\omega\over 2\pi}\left[-\bar{X}_a(-\bar\omega)G^0_R \bar{X}_r(\bar\omega)+{i\over 2}\bar{X}_a(-\bar\omega)G_{sym} \bar{X}_a(\bar\omega)\right],
\end{align}
where $G_{sym}(\bar\omega)=-(1+2n(\bar\omega)) \Im{G^0_R(\bar\omega)}$. From this boundary action we can extract the Langevin equation, see~\cite{Son:2009vu} for details,
\begin{align}
 [-M_0 \bar\omega^2+G_R(\bar\omega)]\bar{X}_r(\bar\omega)=\bar\xi(\bar\omega),\ \ \ \ \vev{\bar\xi(\bar\omega)\bar\xi(\bar\omega)}=G_{sym}(\bar\omega),
\end{align}
with $G_R(\bar\omega)=G^0_R(\bar\omega)+M_0 \bar\omega^2$ and $M_0$ is the zero temperature mass of the external quark in ``bar'' coordinates. Its finite temperature mass is given by
\begin{align}
 M_{T_H}={\gamma\over 2\pi\alpha'}\int_{r_c}^{r_b}dr \sqrt{-g_0},
\end{align}
with $r_c\sim T_H$. For $r_b\gg r_c$. The integral in zero temperature mass formula is dominated by region near the UV-cutoff and so
\begin{align}
 M_0={\gamma\over 2\pi\alpha'}r_b\sqrt{1-v^2}={r_b\over 2\pi\alpha'}.
\end{align}
One can also check using formula in~\cite{Herzog:2006gh} that this is equal to the zero temperature mass of external quark in the original general rotating metric (\ref{general metric}).

Now let's compute retarded green function $G^0_R$ and expand it up to $r_b^0$ using the matching technique~\cite{Harmark:2007jy} which can also be found in~\cite{Atmaja:2010uu} for AdS case. In the low frequency limit, 
\begin{align}
 F^+_{\bar\omega}(r_b)&=C^+ F^+ + C^- F^-,
\end{align}
with
\begin{align}
F^\pm=\left(1-\mp i{\bar\omega\over r_b}\right)e^{\pm i{\bar\omega\over r_b}},\ \ \ C^\pm={1\over 2}\left(1-i b\mp \left[{g_x h_t h_r \over(-g_0)}\right]_{r_c} {1\over \gamma^2 \omega^2} \right),
\end{align}
where $b$ is an arbitrary number.
The result in a leading order of $\bar\omega$ is given by
\begin{align}
 G^0_R(\bar\omega)&=-{r_b\over 2\pi\alpha'} \bar\omega^2-i\left[{g_x h_t h_r \over(-g_0)}\right]_{r_c}{1\over 2\pi\alpha'\gamma^2}\bar\omega \notag \\
&=-{r_b\over 2\pi\alpha'} \bar\omega^2-i{g_x(r_c)\over 2\pi\alpha'\gamma^2}\bar\omega.
\end{align}
One then can extract the Langevin equation written in terms of ``bar'' coordinates as follow:
\begin{align}
 M_0 {\partial^2 \bar{X}(\bar{t}) \over \partial \bar{t}^2}+{g_x(r_c)\over 2\pi\alpha'\gamma^2}{\partial \bar{X}(\bar{t}) \over \partial \bar{t}}={\bar\xi(\bar{t})},\ \ \ \ \vev{\bar\xi(\bar{t})\bar\xi(\bar{t}')}=2\bar{T}{g_x(r_c)\over 2\pi\alpha'\gamma^2}\delta(\bar{t}-\bar{t}').
\end{align}
Writing in term of time coordinate $t$ and using that $\bar{X}(\bar{t})=\gamma X(t)$, we obtain
\begin{align}
 \gamma^3 M_0 {\partial^2 X(t) \over \partial t^2}+{g_x(r_c)\over 2\pi\alpha'}{\partial X(t) \over \partial t}={\xi(t)},\ \ \ \ \vev{\xi(t)\xi(t')}=2T{g_x(r_c)\over 2\pi\alpha'}\delta(t-t'),
\end{align}
where we have defined $\xi(t)=\bar\xi(\bar{t})$. Here once again we show that for general metric (\ref{general metric}) the mass of Brownian particle is given not by the zero temperature mass of the external quark but rather its effective mass $M_{eff}\equiv\gamma^3 M_0>M_0$, for $v>0$.

\section{Discussion and Conclusion}

The result in (\ref{BTZ result}) justifies the use of the standard linear Langevin equation for the co-rotating case of Brownian motion on BTZ black hole in the calculation of~\cite{Atmaja:2012jg}. It also implicitly justifies the random-random force correlator approach in the aforementioned paper. The random-random force correlator is equal to the one calculated in the approach of~\cite{Atmaja:2012jg} which indicates that the external quark behaves as a Brownian particle. Since there is no mass dependent in its explicit formula, or in the same meaning that the friction coefficient is inverse to the mass, the mass in the Langevin equation is not necessary the zero temperature mass of external quark, $M_0$, in fact the Brownian mass should be given by an effective mass, $\gamma^3 M_0$. It is an interesting fact that an external quark moves co-rotatingly with the fluid has larger mass compare to its relativistic mass, $M_{rel}=\gamma M_0$. It might gives an indication that an object moves under influence of a black hole will have a larger mass than its relativistic mass. This was observed in the longitudinal fluctuation on a moving string under five dimensional AdS-Schwarzschild black hole~\cite{Giecold:2009cg}\footnote{In~\cite{Giecold:2009cg}, the relative factor between the effective (longitudinal) mass and the relativistic (tranverse) mass can be traced back from the relation between energy and transverse momentum of an emitted parton.}. In a different context this $\gamma^3$ factor also appeared in four dimensional the black hole mass ration between the AdS-Schwarzschild black hole and the Kerr-AdS black hole~\cite{Atmaja:2011ji}.

The effective mass of co-rotating external quark in more than three dimensional, up to five dimensional, rotating black holes also shows a generic behaviour, as in the BTZ black hole case, although we limit our discussion to the fluctuation along the longitudinal direction with a general metric (\ref{general metric}). This metric is asymptotically $AdS_3$ and is a subspacetime of the original metric for some fixed (compact) coordinates. For example, the general metric (\ref{general metric}) can be obtained from four dimensional Kerr-AdS black hole by taking the equatorial plane, $\hat\theta=\pi/2$, in asymptotically ``cannonical'' AdS coordinate. One may try to take a non-equatorial plane, $\hat\theta\neq\pi/2$, however the presence of anisotropic force will pull the external quark to the equatorial plane~\cite{NataAtmaja:2010hd}. Hence the stationary string solution is only possible at the equatorial plane. However for five dimensional Kerr-AdS black hole and topological KMV-AdS black hole~\cite{Klemm:1997ea}, we have not found in the literature the study on anisotropic drag force that leads to stationary string solution on some fixed compact coordinate-dependent. Extending the calculation to more than five dimensional rotating black holes is more involved since it contains more than one compact coordinate-dependent. Nevertheless, we may still use general metric (\ref{general metric}) requiring only asymptotically $AdS_3$ and string fluctuation along longitudinal direction of stationary string solution $X$.

\section*{Acknowledgments}
This work is supported by Kompetitif-LIPI grant 2013 under the project of ``Dinamika Fluida pada Energi Tinggi''.

\end{document}